\documentclass[sigconf]{acmart}
\AtBeginDocument{%
  }

\setcopyright{acmlicensed}
\copyrightyear{2025}
\acmYear{2025}
\acmConference[]{CHI'25 Workshop: Everyday AR through AI-in-the-Loop}{April 2025}{New York, NY, USA}

\begin{document}

\title{Augmenting Human Cognition through Everyday AR}


\author{Xiaoan Liu}
\email{xiaoanliu@nyu.edu}
\affiliation{%
  \institution{New York University}
  \city{New York}
  \state{NY}
  \country{USA}
}

\renewcommand{\shortauthors}{Liu et al.}

\begin{abstract}
As spatial computing and multimodal LLMs mature, AR is tending to become an intuitive "thinking tool," embedding semantic and context-aware intelligence directly into everyday environments. This paper explores how always-on AR can seamlessly bridge digital cognition and physical affordances, enabling proactive, context-sensitive interactions that enhance human task performance and understanding.
\end{abstract}






\maketitle

\section{INTRODUCTION}
A persistent theme in technology evolution is that once a foundational layer matures and becomes commoditized, the real competition moves to the "interface layer." Historically, we have seen this in the shift from mainframe computing to personal computers, then from web browsers to mobile devices. At each transition, an enabling technology becomes sufficiently widespread and affordable, opening space for a new, more intuitive interface.

We are now witnessing a similar pivot in the age of spatial computing and multimodal large language models (MLLMs). As head-worn devices shrink and edge connectivity matures, Augmented Reality stands to become the predominant "interface layer" that seamlessly merges the digital realm with everyday physical contexts. Paired with AI that can interpret speech, gaze, gestures, and environmental signals, AR can go beyond overlaying disjointed information to become a cognitive amplifier—a "thinking tool for reality." \cite{Horvitz1999} 
\section{REALITY AS THE CANVAS}

In this shifting paradigm, augmented reality moves from being a mere overlay of graphics to an immersive layer of intelligence woven seamlessly into our daily environment.

\subsection{AR as an Always-On Knowledge Layer}
In a smartphone-centric world, we measure "engagement" by screen time: the number of minutes an app occupies in the foreground. But with always-on AR glasses, attention transforms. In AR, the user might glance at an object, utter a partial phrase, or make a subtle gesture—and instantly the system offers assistance. Our "context sharing"—the exchange of situational information between user and system—shifts from traditional apps into an \emph{ambient, environment-embedded} dialogue. \cite{Ishii1997, McGuffin2019}

For example, instead of having to switch to a separate translation app, you can simply look at text through your AR glasses, and the system will automatically translate or summarize it for you. If you say something like "I need more info on this brand" while viewing a product label, the AI in your glasses quickly retrieves the relevant information.

\subsection{Moving from "Screens" to "Scenes}
The fundamental difference is that we are no longer locked into 2D windows. AR can unify your physical space into a single interactive "scene" where every wall, tabletop, or building façade becomes a potential surface for dynamic content. Our human attention is guided not by explicit screen boundaries, but by the inherent salience of physical and social cues.

This shift hints at deeper cognitive augmentation: the device can unobtrusively sense your context, reduce superfluous mental overhead, and highlight the most relevant information or actions at the exact moment you need them.

\begin{figure}[h]
    \centering
    \includegraphics[width=0.45\textwidth]{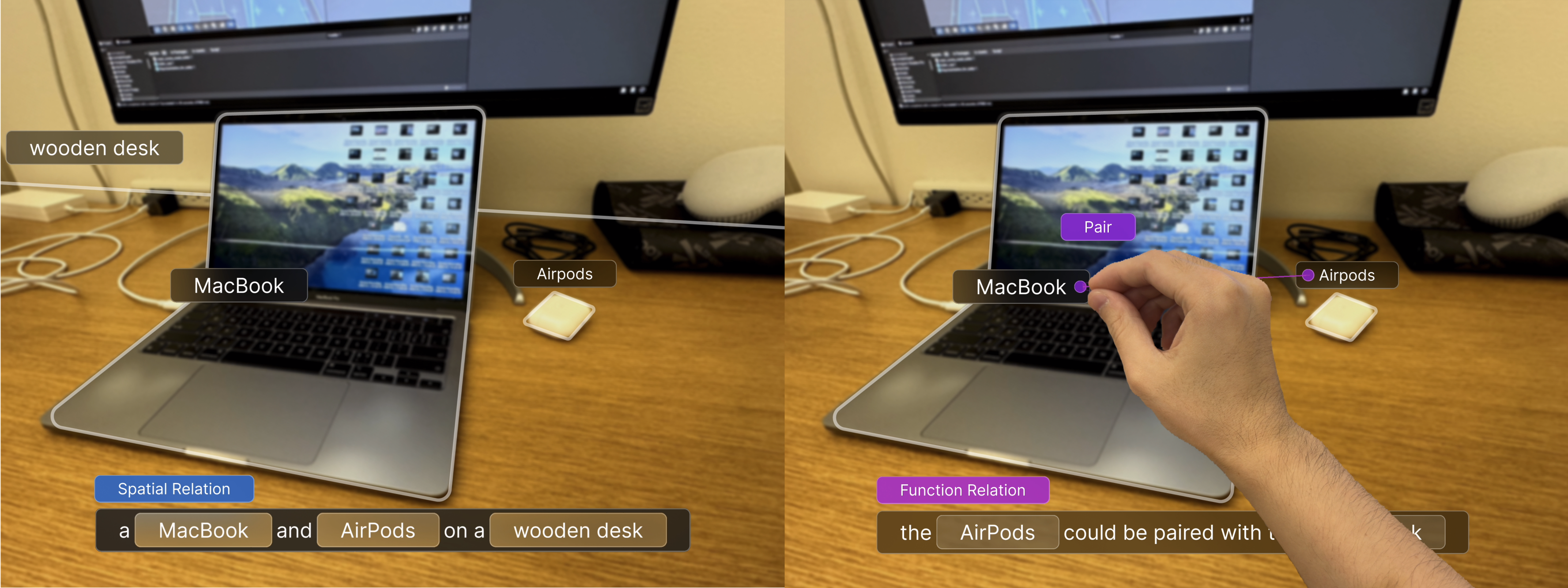}
    \caption{A scenario in AR where a MacBook and AirPods on a desk are labeled, with a 'Pair' prompt suggesting their connection, allowing users to initiate pairing with a gesture by directly manipulating the virtual connector.}
    \label{fig:macbook}
\end{figure}

Figure~\ref{fig:macbook} demonstrates a basic example of how \emph{always-on knowledge} can facilitate immediate insights about physical objects. Instead of manually searching for Bluetooth settings, a quick gesture or voice command in context triggers a pairing workflow.

\section{THE SEMANTIC LAYER FOR REALITY}
Reality computing involves embedding a semantic layer over the real world. This is not just about labeling objects or overlaying text; it is about endowing the AR system with an interpretive, context-sensitive understanding that lets it act as an extension of the user's cognition.

\subsection{Natural and Contextual Interaction}
By combining the user's gaze vector with voice or gesture input, AR systems can infer user intent even if commands are ambiguous. For instance, users might say "Change that," while glancing at a lamp, and the system interprets that they want to dim the light. Large language models—upgraded to handle images, 3D scans, or sensor data—can identify objects, interpret vague pronouns, track user references, and handle natural conversation. They become the semantic "bridge" tying real-world context to digital intelligence. \cite{Endert2012, Kelly2018}

\subsection{Proactive Thinking Tools}
\begin{figure}[h]
    \centering
    \includegraphics[width=0.45\textwidth]{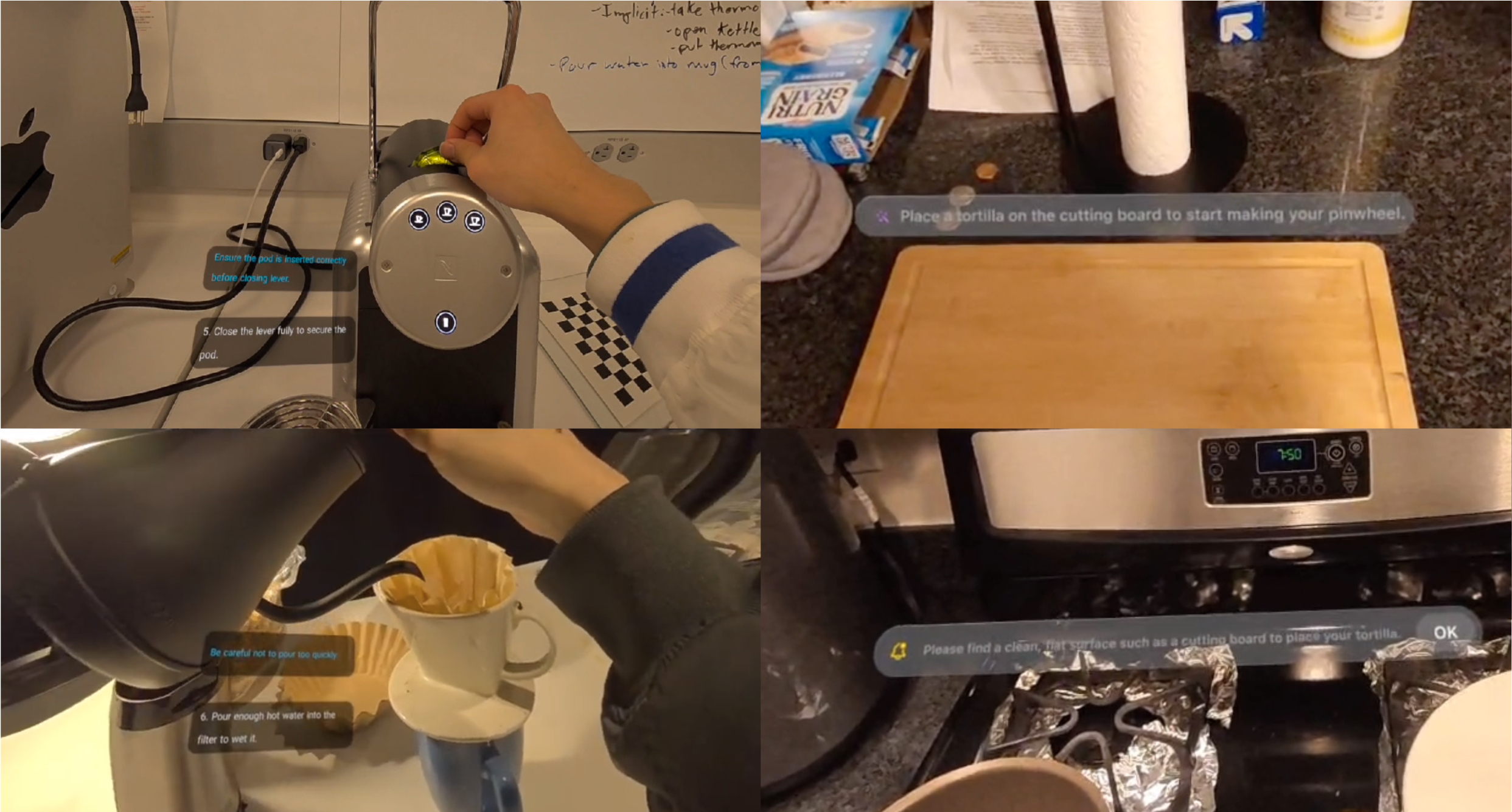}
    \caption{RealiTips providing step-by-step guidance in AR.}
    \label{fig:realiTips}
\end{figure}

Adaptive guidance involves using an AR system to proactively walk the user through step-by-step tasks. Instead of requiring manual step-switching or referencing static instructions, the system "sees" what the user is doing in real time and autonomously advances through a tutorial. Figure~\ref{fig:realiTips} illustrates this idea with our "RealiTips" prototype.

What distinguishes RealiTips from traditional tutorials is that it advances instructions based on actual user progress, creating a more natural learning flow. Predictive suggestions further reduce mental overhead by anticipating user needs. For instance, if a user sets a half-assembled IKEA shelf on the floor, the system might proactively highlight the needed screwdriver and the correct set of screws.

\begin{figure}[h]
    \centering
    \includegraphics[width=0.45\textwidth]{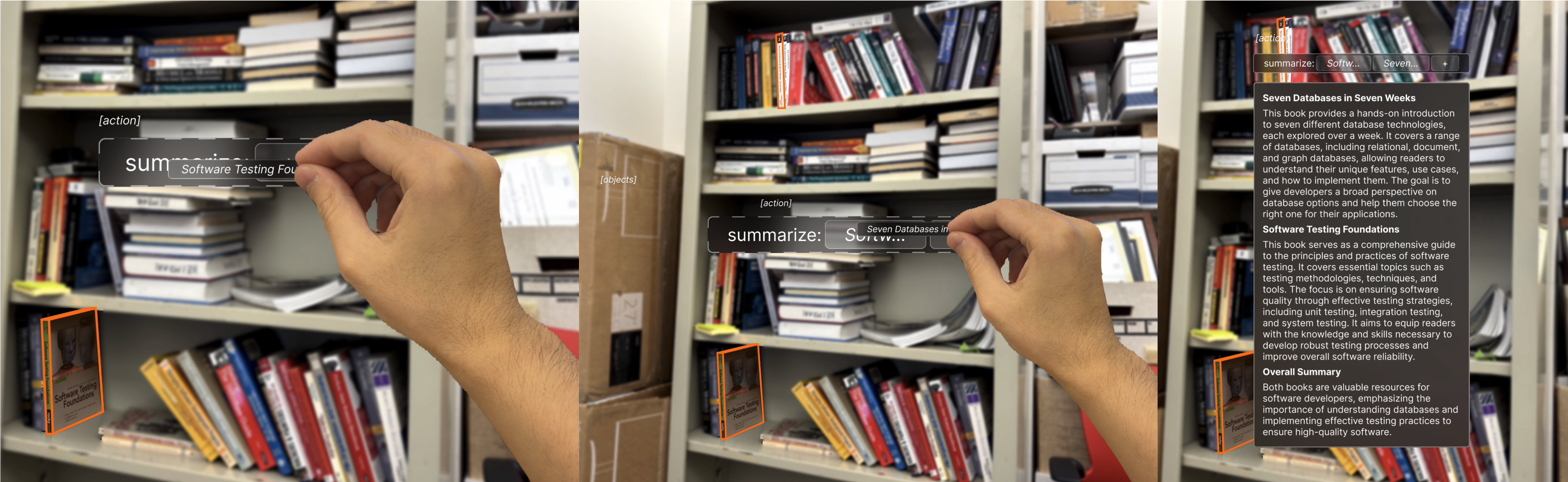}
    \caption{A library scenario showing AI-assisted semantic search. The AR system recognizes multiple books, such as \emph{Software Testing Foundations} and \emph{Seven Databases in Seven Weeks}, allowing the user to ask for summaries, related topics, or best practices in context.}
    \label{fig:library}
\end{figure}

Figure~\ref{fig:library} offers an example of combining \emph{contextual interaction} and \emph{semantic augmentation} in a library. The user can hover a finger over a book's bounding box (e.g., \emph{Software Testing Foundations}) and verbally request "summarize," "about the author," or "related books." The system harnesses large language models to provide concise text overlays or spoken feedback. This transforms a physical shelf-browsing experience into a context-rich knowledge discovery process.

\section{LEVERAGING THE AFFORDANCES OF THE PHYSICAL WORLD}
Heads-up AR displays uniquely exploit the "affordances" of the physical world \cite{gibson2014theory}. By presenting spatially anchored information directly in one's field of view, these systems overlay rich digital context onto real-world cues (e.g., a handle that invites grasp, a knob that invites rotation), effectively scaffolding user cognition. Instead of memorizing sequences or referencing an external screen, the user can rely on direct, in-place guidance. 

\begin{figure}[h]
    \centering
    \includegraphics[width=0.45\textwidth]{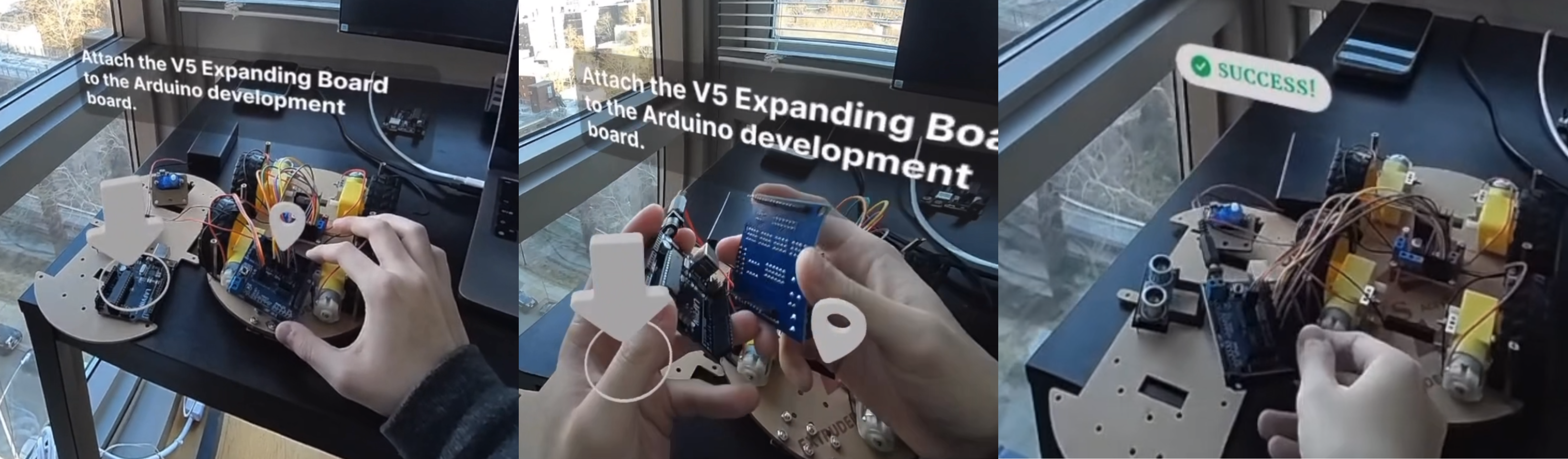}
    \caption{Step-by-step instructions for assembling an Arduino-based robotic platform. The AR system tracks the user's progress and displays targeted prompts (e.g., ``Attach the V5 Expanding Board''), reducing errors and accelerating learning.}
    \label{fig:arduino}
\end{figure}

Figure~\ref{fig:arduino} shows a prototypical scenario where a user assembles a small robotic platform with an Arduino. Each assembly step appears over the relevant part, with arrows, labels, or textual prompts. The system detects partial completion (e.g., whether screws are attached), updating the user's progress in real time. By integrating these instructions into the user's field of view and mapping them to physical objects, AR reduces the cognitive effort needed to understand or recall tasks, allowing users to better focus on the activity at hand. \cite{Monteiro2023, Zhu2022}

This real-time interplay between the physical affordances of reality and contextual AI feedback not only accelerates task performance but also deepens understanding, as users learn how objects and actions connect in both physical and digital realms. The environment itself becomes a "living manual," where guidance and knowledge are actively anchored to items and surfaces, freeing users from distracting interface layers while still providing moment-to-moment assistance.

\section{CONCLUSION}
AR with MLLMs transcends being a mere interface, evolving into a potent "thinking tool" that bridges physical tasks and digital knowledge. From pairing headphones to exploring books and guiding cooking or robot assembly, reality computing seamlessly integrates with daily cognition. Ultimately, by embedding intelligence into our physical world, AR will enhance our understanding and extend cognitive capabilities, once the realm of science fiction.

% \begin{acks}
% \end{acks}

\bibliographystyle{ACM-Reference-Format}
\bibliography{sample-base}

\appendix









\end{document}